\documentclass[a3paper,12pt]{article}

\usepackage[utf8]{inputenc}
\usepackage{graphicx}
\usepackage{amsxtra}
\usepackage{url}

\usepackage{latexsym}
\usepackage{amssymb}

\usepackage{ulem}

\newcommand{\ser}{S\'ersic}

\def\bc{\begin{center}}
\def\ec{\end{center}}

\def\be{\begin{equation}}
\def\ee{\end{equation}}
\def\beq{\begin{eqnarray}}
\def\eeq{\end{eqnarray}}
\def\bfig{\begin{figure}}
\def\efig{\end{figure}}
\def\bnum{\begin{enumerate}}
\def\enum{\end{enumerate}}

\begin{document}

\vspace{1cm}
\bc
{\Large {\bf Bulges and discs of spiral galaxies: edge-on perspective}} \\
\vspace{0.5cm}
{N.YA.~SOTNIKOVA, V.P.~RESHETNIKOV and A.V.~MOSENKOV}\\
\vspace{0.5cm}
{St.Petersburg State University, St.Petersburg 198504, Russia}\\
\vspace{0.5cm}
{\it Received 1 December 2010}
\ec

\label{firstpage}
\begin{abstract}

We present a sample of edge-on spiral galaxies both of early
and late types. The sample consists of 175 galaxies in the $K_s$-filter, 
169 galaxies in the $H$-filter and 165 galaxies in the $J$-filter. 
Bulge and disc decompositions of each galaxy image, taken from the 
Two Micron All Sky Survey (2MASS), were performed. We discuss several scaling 
relations for bulges and discs which indicate a tight link between their 
formation and evolution. We show that galaxies with bulges fitted by the 
\ser\ index $n < 2$ (pseudobulges) have quite different distributions 
of their structural parameters than galaxies with $n \geqslant 2$ bulges 
(classical bulges). First of all, the distribution of the apparent bulge axis 
ratio $q_\mathrm{b}$ for the subsample with $n < 2$ can be attributed 
to triaxial, nearly prolate bulges, while $n \geqslant 2$ bulges seem to be 
oblate spheroids with moderate flattening. Secondly, the Photometric Plane of 
the sample bulges is not flat and has a prominent curvature towards small 
values of $n$. Thirdly, despite of the existence of a clear relation between 
the flattening of stellar discs $h/z_0$ and the relative mass of a spherical 
component, the distributions over both parameters are quite different for 
galaxies possesing classical bulges and pseudobulges.

\end{abstract}

{\it Keywords}:
Galaxies: infrared --- photometry --- spiral edge-on galaxies --- 
fundamental parameters.

\section{Introduction}

In recent decades a number of studies concerning structural properties of 
galaxies has been performed. The results of decompositions of near face-on 
disc dominated galaxies as well as of early-type galaxies yeilded a lot of 
correlations between structural parameters of discs and bulges 
\cite{dej, mol2, gra01, cour96}.

Among all galaxies edge-on ones are of great interest because they provide 
a unique possibility to obtain information about the vertical structure of 
a disc and a bulge. The disc scaleheight $z_0$ together with the disc 
scalelength $h$ determine the relative thickness of a stellar disc for each 
galaxy. This ratio as well as the bulge flattening $q_\mathrm{b}$ (the 
ratio of isophote semi-axes) may put constraints on the disc and bulge 
formation processes and their secular evolution.

There are some difficulties in studying edge-on galaxies 
and obtaining their structural parameters because of the dust 
extinction. The possible solution is to analyse observations in the NIR 
passbands where the extinction is drastically reduced. A detailed study of 
the discs of edge-on galaxies is described in many papers 
\cite{degr1, dal02, biz, kre02, kre05, zas02, biz10}. 
Almost all these papers dealt with bulgeless or late types 
edge-on galaxies and did not focus on the properties of bulges. 
Our study is intended to investigate structural parameters of edge-on 
galaxies both of early and late types and to join analysis of their bulges 
and discs in the near-infrared bands. 

For our purpose we used the 2MASS-selected Flat Galaxy Catalog (2MFGC) 
\cite{mit} as a source of objects and the 2MASS survey 
\cite{skr} as a source of IR image data. We also used the Revised Flat 
Galaxies Catalog (RFGC) \cite{kar99} to add late-type galaxies to our sample. 
Our celection criteria are presented in \cite{mos}.
The absolute numbers of galaxies of various morphological types (taken from 
the LEDA) and their percentages are listed in the Table~\ref{Types}. 

\begin{table}
 \centering
 {\caption{Morphological types of the sample galaxies and
their fractions} \label{Types}} 
 \begin{tabular}{ccc||ccc}
 \hline 
 \hline
 Type     & Number & Percentages  & Type & Number & Percentages\\ \hline
 S0, S0-a & 41 & 23.4 & Sbc   & 21  & 12.0 \tabularnewline
 Sa       & 11 & 6.3  & Sc    & 36  & 20.5 \tabularnewline
 Sab      & 15 & 8.6  & Scd   & 4   & 2.3  \tabularnewline
 Sb       & 47 & 26.9 & {\bf Total} & {\bf 175} & {\bf 100}  \tabularnewline
 \hline
\end{tabular}
\end{table}

The distances of the galaxies are taken from NASA/IPAC Extragalactic
Database (NED) with the Hubble constant $H_0=73$ km s$^{-1}$/Mpc, 
$\Omega_\mathrm{{matter}}=0.27$, and $\Omega_{\mathrm{vacuum}}=0.73$). 
To investigate some kinematical and dynamical properties of the sample
galaxies we added information about maximum rotation velocity 
for each galaxy if it was presented in the LEDA data base. The sample 
consists of 175 galaxies in the $K_s$-filter, 169 galaxies in the $H$-filter 
and 165 galaxies in the $J$-filter. The full list of the sample galaxies 
with the basic information about them can be found in \cite{mos}.

According to the $V/V_\mathrm{max}$ test \cite{schm} our full sample is 
incomplete but the subsample of galaxies with angular radius 
$r\geqslant 60''$ is complete ($V/V_\mathrm{max}=0.49 \pm 0.03$). 
This subsample consists of 92 galaxies (47 early-type galaxies and 
45 late-type galaxies). 

\section[]{Two-dimensional bulge/disc decomposition}

We used BUDDA (Bulge/Disc Decomposition Analysis) v2.1 code \cite{des}
to perform two-dimensional decomposition of galactic images 
taken from the database of the All-Sky Release Survey Atlas.

The photometric model consists of two major components: a bulge and a disc. 
The disc is assumed to be axisymmetric, transparent and is represented by 
an exponential distribution of the luminosity density. It can be 
described by the face-on central surface brightness 
$S_\mathrm{0,d}$ (in mag per arcsec$^2$), the scalelength $h$ and the 
`isothermal' scaleheight $z_0$.

For the bulge surface brightness profile the \ser\ law is adopted 
\cite{ser}. The main parameters of this profile are: the central effective 
surface brightness $\mu_\mathrm{0,b}$ expressed in mag per arcsec$^2$, 
the effective radius of the bulge $r_\mathrm{e,b}$ and the \ser\ index $n$, 
defining the shape of the profile. We also assume that bulge isophotes 
are described by ellipses with the flatness $q_\mathrm{b} = b/a$, 
where $b$ is the semi-minor axis and $a$ is the semi-major axis of an 
isophote. The \ser\ index of $n=4$ represents the de Vaucouleurs profile 
which is very popular to describe the surface brightness distribution of 
bright elliptical galaxies and of bulges in early-type spirals. 
The \ser\ index of $n=1$ represents the exponential 
profile of bulges in late-type spirals, of galactic discs and of dwarf 
elliptical galaxies (see e.g. \cite{dev48, dev53, fre, gra01}).

To check the reliability of our 2D decompositions, we compared our 
results with the results obtained by other authors. In the $K_s$ filter, 
our sample has 30 galaxies in common with the sample in \cite{biz} 
and 14 galaxies that have been investigated in \cite{degr1}. 
All mentioned authors used different decomposition procedures but the 
results of comparison are in good agreement \cite{mos}.

\section[]{New evidences for different origin of bulges and 
pseudobulges}

Statistical correlations between structural parameters give us a clue for 
understanding the formation and evolution of spiral galaxies. Here we present 
several correlations of the main parameters of the galaxies and some 
new evidences in favour of existence of two quite different populations of 
bulges in spiral galaxies --- classical bulges and pseudobulges.

For edge-on galaxies morphological types are very subjective because the 
spiral arms are not seen in this case. That is why we distinguish 
the galaxies in our sample using the \ser\ index $n$ because there 
is a good correlation between $n$ and the ratio of bulge and disc 
luminosities $B/D$ that is known to be a key parameter for classifying spiral 
galaxies on the Hubble sequence \cite{dej, gra01}

\begin{center}
 \begin{figure*}
 \begin{center}
  \includegraphics[width=5.4cm, angle=-90]{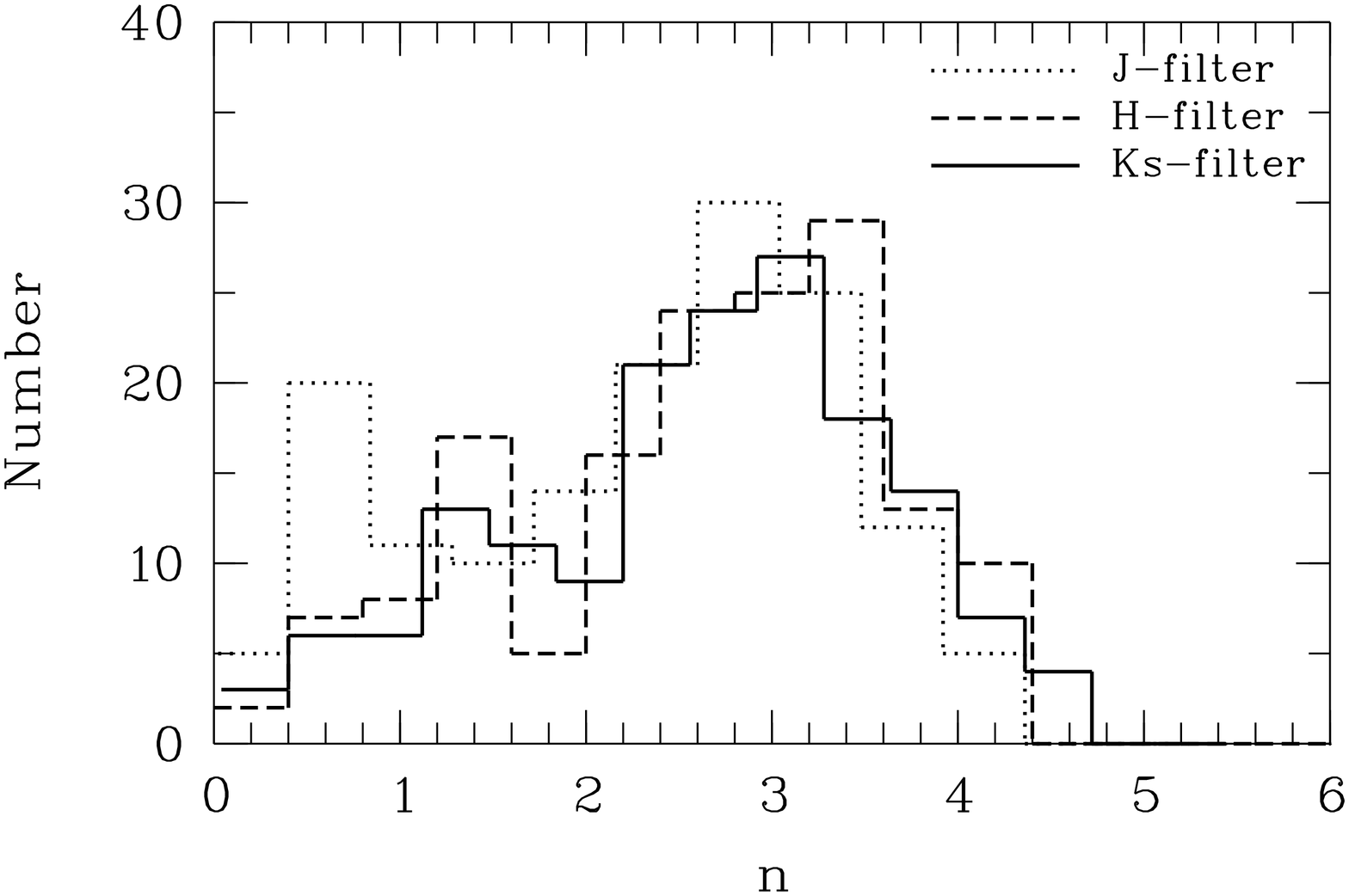}
  \includegraphics[width=5.4cm, angle=-90]{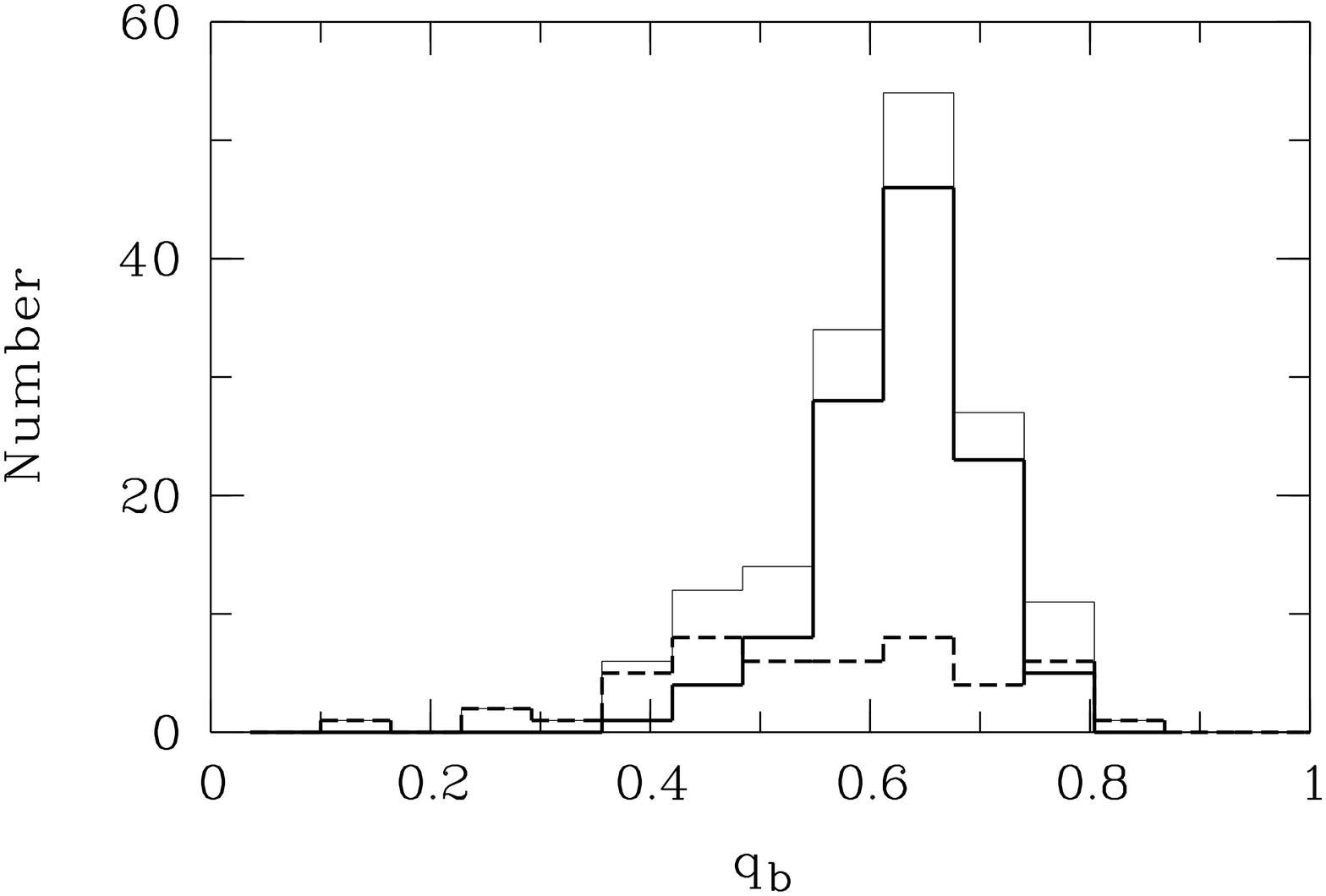}
 \end{center}
 \caption{\small{Distributions of the sample galaxies over 
 the \ser\ index of bulges (left) and the model bulge axis ratio 
 $q_\mathrm{b}$ in the $K_s$-band (right). At the right panel 
 the thin line indicates the distribution for the whole sample, 
 the solid line  corresponds to the subsample of bulges with 
 $n \geqslant 2$ and the dashed line shows the distribution for 
 the subsample of bulges with $n < 2$.}}
\label{Distrib}
\end{figure*}
\end{center}

The distributions over the \ser\ indices of galaxies in three passbands 
are shown in Fig.~\ref{Distrib} (the left plot). The distributions 
demonstrate a weak bimodality which may reflect the existence of two families 
of bulges: bulges with $n \geqslant 2$ and bulges with $n < 2$. This 
bimodality is known for some galaxy samples (e.g. \cite{fis07}) and probably 
is real for our sample. As was reported earlier (see e.g. \cite{fis07,fis08}) 
such a bimodality correlates with morphological type of the bulge. 
The classical bulges have $n \geqslant 2$ and the so-called pseudobulges 
have $n < 2$. The classical bulges appear to be similar to elliptical 
galaxies and have similar properties. It is assumed that these systems were 
built via minor and major merging. The galaxies with the pseudobulges are 
thought to be formed via disc instabilities and secular evolution and have 
disc-like apparent flattening. 

\subsection[]{The intrincic shape of bulges}

In Fig.~\ref{Distrib} (the right plot) we demonstrate the distributions of 
model bulge axis ratio $q_\mathrm{b}$ for edge-on view. In the case of 
edge-on galaxies this parameter describes directly the bulge flatness and the 
intrinsic 3D structure of bulges if they are assumed to be oblate spheroids. 
The median value of this parameter is $\sim 0.63$, 
independently of the band. This is in good agreement with 
\cite{mor98, noo}. Thus, in general bulges are definitely nonspherical 
and flattened. However the question about 3D shapes of bulges and their 
possible triaxiality are still open, despite of the fact that its 
solution is thought to be crucial for testing different scenarios of galaxy 
formation.

Our sample of edge-on galaxies allows us to distinguish clearly 
the difference in the bulge equatorial ellipticities for galaxies with 
classical bulges and pseudobulges.
As one can see in Fig.~\ref{Distrib} (the right plot) for classical bulges 
the distribution of $q_\mathrm{b}$ has a rather narrow peak at 
$q_\mathrm{b} \approx 0.65$. This may reflect the fact that the bulges 
are nearly oblate spheroids with moderate flattening. The 
distribution over $q_\mathrm{b}$ for pseudobulges is very wide, 
spreading from very flat bulges up to nearly spherical ones. Such a 
distribution may be attributed to definitely triaxial, near prolate bulges 
that are seen from different projections --- along the major axis and 
perpendicular to it. But we can not exclude that triaxial shape of bulges 
in late-type galaxies may hide the presence of bars that went thick in 
the vertical direction during secular evolution. 

Our findings conradict the results obtained in \cite{men2}. These 
authors recovered the PDF for the sample of 148 unbarred S0-Sb 
galaxies at intermediate inclination angles and concluded that bulges 
with $n < 2$ show a larger fraction of oblate axisymmetric bulges, 
a smaller fraction of triaxial bulges, and fewer prolate axisymmetric 
bulges with respect to bulges with $n \geqslant 2$.

\subsection[]{Photometric Plane and \ser\ index}

\begin{center}
 \begin{figure*}
 \begin{center}
  \includegraphics[width=6.0cm,  angle=-90]{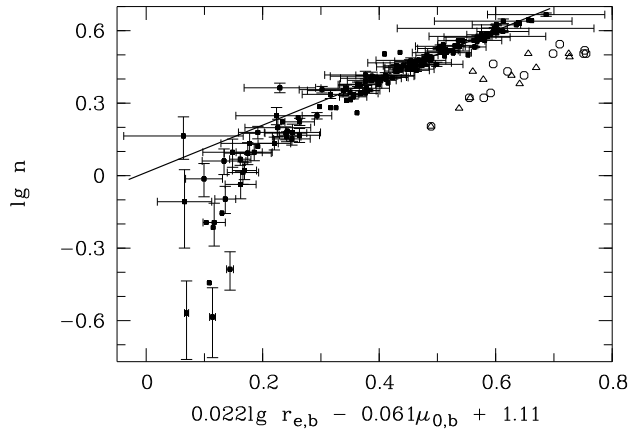}
 \end{center}
 \caption{\small{
 The Photometric Plane  built for bulges with $\lg n \geqslant 0.2$ of the 
 complete subsample in the $K_s$-band. The points corresponding to bulges 
 with $\lg n \leqslant 0.1$ do  not lie in this plane. The model merger 
 remnants of disc galaxies are shown with open circles and triangles 
 \cite{ace}. The model data are shifted by a constant along the horizontal 
 axis.}}
 \label{BulgeCorr1}
 \end{figure*}
\end{center}

The difference between bulges and pseudobulges arises while 
building the Photometric Plane (PhP) for them. This plane represents a tight 
correlation of the \ser\ index $n$ with the central surface brightness 
$\mu_\mathrm{0,b}$ and the effective radius of a bulge $r_\mathrm{e,b}$ 
\cite{kho1, kho2}. Performing the least-squared fit of an expession 
$\lg n = a \lg r_\mathrm{e} + b \, \mu_\mathrm{0,b} + c$ we obtained a 
surprising result. It occurred that the galaxies with $\lg n \geqslant 0.2$ 
($n \geqslant 1.58$) lying on the PhP, that is built for these galaxies, 
has a small scatter, but the galaxies with $\lg n \leqslant 0.1$ 
($n \leqslant 1.26$) do not correspond to this plane and form their own plane 
(Fig.~\ref{BulgeCorr1}). The main difference between our sample and the samples 
of the above-mentioned works is the range of the shape parameters $n$. 
Our sample contains a substantial amount of bulges with $\lg n \leqslant 0.1$. 
For the sample in \cite{kho1} the low boundary for $n$ lies at 
$\lg n \geqslant 0.2$, that is why the curvature of the Photometric Plane 
towards small values of $n$ for bulges was not noticed earlier. 

It is not clear whether two distinct parts of the curve in 
Fig.~\ref{BulgeCorr1} reflect the different origin of bulges lying in 
the corresponding regions. However, the collisionless merger remnants of 
disc galaxies obtained in $N$-body simulations and fitted by a \ser\ profile 
fairly well reproduce the slope of the PhP in the region of 
$\lg n \geqslant 0.2$ \cite{ace}. We put on the data from \cite{ace} in our 
Fig.~\ref{BulgeCorr1} and found the numerical data to be consistent with our 
observational ones. The curvature of the PhP towards small values of $n$ may 
reflect the quite different nature of such bulges, formed, for example, via 
secular evolution of discs.

\begin{center}
 \begin{figure*}
 \begin{center}
  \includegraphics[width=5.4cm,  angle=-90]{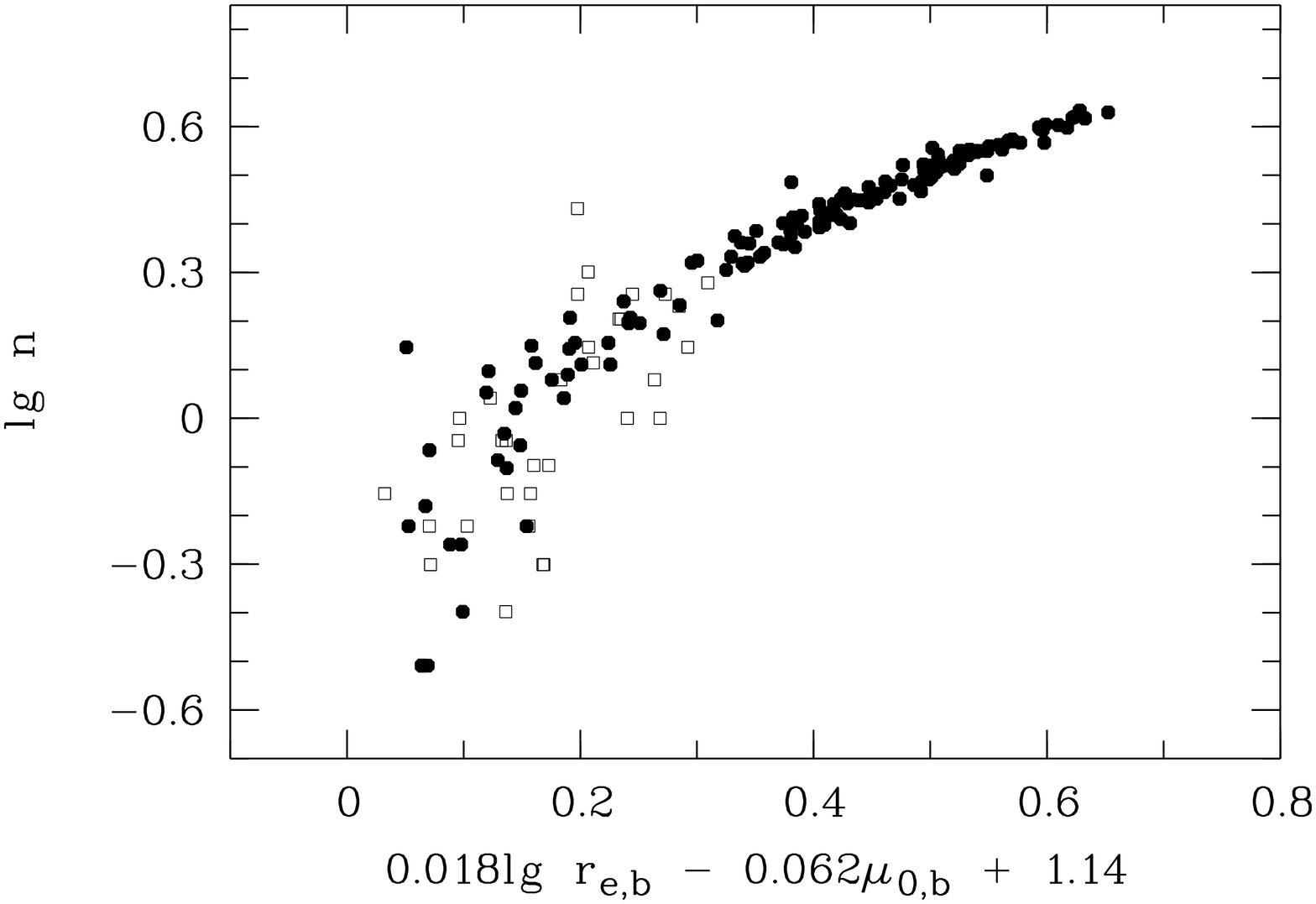}
  \includegraphics[width=5.4cm,  angle=-90]{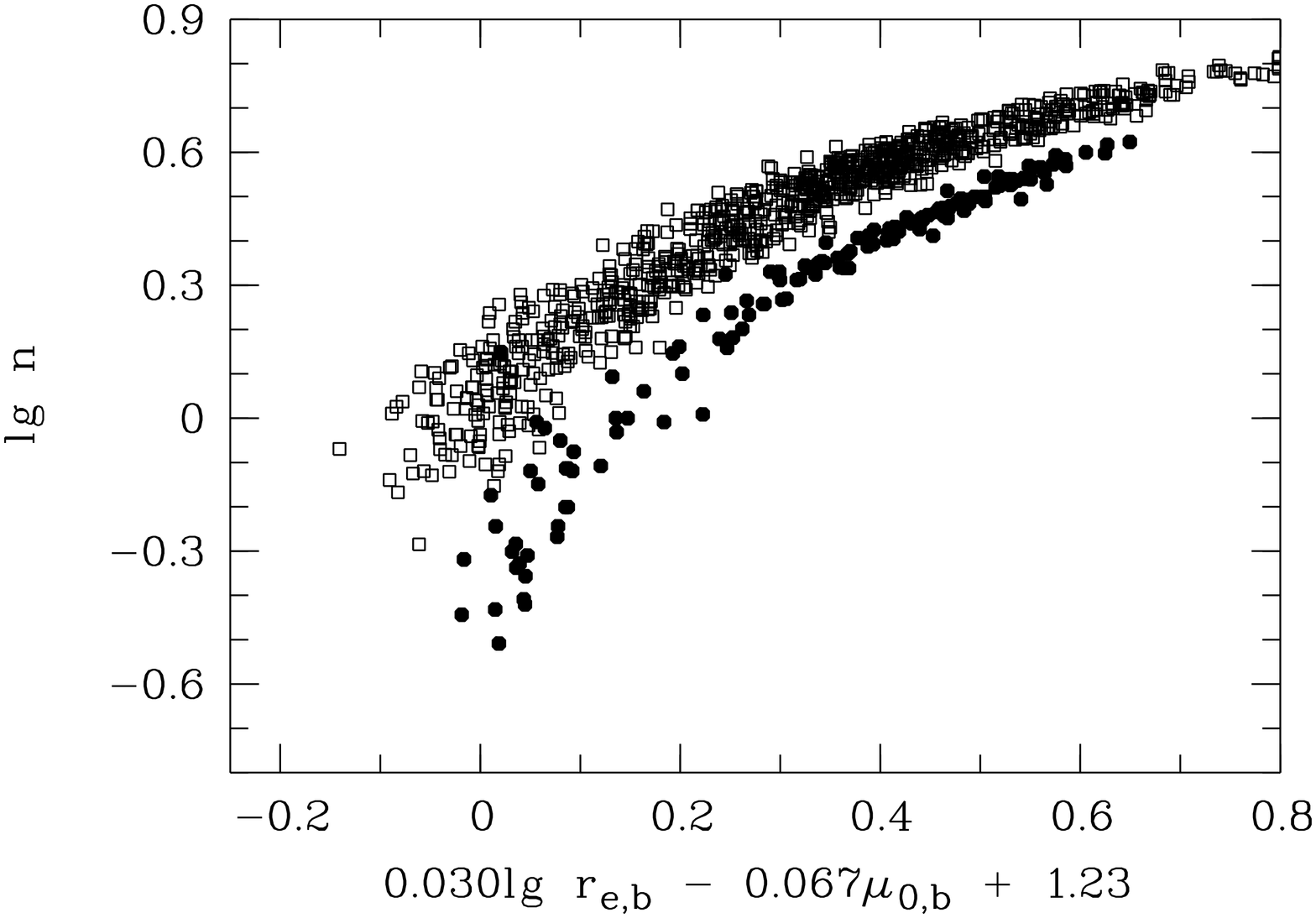}
 \end{center}
 \caption{\small{
 The Photometric Plane built for bulges with $\lg n \geqslant 0.2$. Left --- 
 the data from \cite{maca} (open squares) in comparison with our data 
 (filled circles). Both data are related to the $H$-band. Right --- 
 the data from \cite{gad09} (open squares, the $i$-band) in comparison 
 with our data (filled circles, the $J$-band).}}
 \label{BulgeCorr2}
 \end{figure*}
\end{center}

The found curvature is not the feature of our decomposition and the 
presence of a large number of galaxies with small angular radii which 
may give a confusing decomposition. We constructed the PhP only for 
galaxies from the complete sample, possesing systems with 
$r \geqslant 60"$. 
The curvature was not noticed earlier because of a lack of substantial 
number of late-type galaxies in the samples that were studied. We constructed 
the Photometric Plane based on the samples 
presented in \cite{maca, gad09}. The first sample contains 121 face-on 
and moderately inclined late-type spirals with derived structural 
parameters of bulges and discs in the $H$-band. Fig.~\ref{BulgeCorr2} 
(the left panel) shows the Photometric Plane for the bulges from 
\cite{maca} together with our data. The plot displays the same feature 
as Fig.~\ref{BulgeCorr1}: a very tight correlation for classical bulges 
and a fairy large scatter of points for pseudobulges. Surprisingly, 
both samples follow the same relation in the region of small values of $n$. 
The sample, presented in \cite{gad09}, consists of around 1000 galaxies with 
determined photometric parameters in the $i$-band. The most close passband 
in our sample is the $J$-band. We display the best fitting relation for our 
data in Fig.~\ref{BulgeCorr2} (right) and added the point from the 
sample \cite{gad09}. There is a shift between two relations which is 
due to the difference in the passbands. But both samples show a 
prominent curvature towards small values of $n$.


\subsection[]{The dark halo and disc flattening}
\label{DHalo}

The disc flattening can be measured directly only for edge-on galaxies. It 
is commonly expressed as the ratio of disc scaleheight $z_0$ to disc 
scalelength $h$. This ratio shows a weak trend with Hubble type 
\cite{degr1}. Zasov et al. \cite{zas02} were the first to found that the disc 
flattening shows a definite trend with the ratio of dynamical mass 
$\mathrm{M}_\mathrm{tot}$ to the disc luminosity provided the dynamical mass 
is defined as the total mass enclosed within the sphere of radius that is 
equal to four disc scalelengths. These authors suggested that discs are 
marginally stable against the growth of perturbations in their planes and 
bending perturbations and implied the linear relation between the ratio 
$h/z_0$ and the ratio of dynamical mass (or dark halo mass) to disc 
luminosity. It is valid at least for bulgeless galaxies. 

The found correlation means that relatively thinner discs in bulgeless 
galaxies tend to be embedded in more massive dark 
haloes. Zasov et al. \cite{zas02} verified the linear trend, obtained 
analytically, for two different samples of edge-on bulgeless 
galaxies with known structural parameters in the $R$ and $K_s$ bands,  
using the HI line width $W_{50}$ as a measure of dynamical mass. For 
our sample, which contains a lot of galaxies with massive bulges, we 
performed the same analysis. 

\begin{center}
 \begin{figure*}
 \begin{center}
  \includegraphics[width=6.0cm, angle=-90]{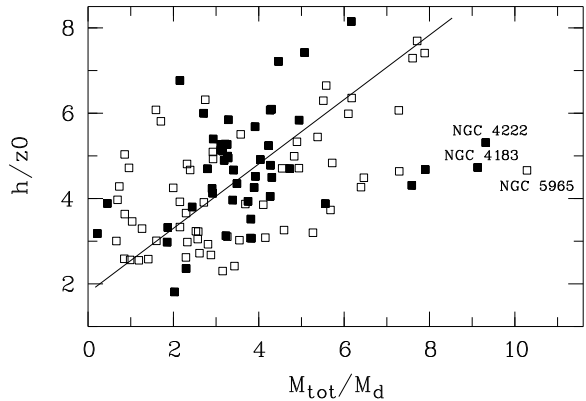}
 \end{center}
 \caption{\small{
 The ratio of $h/z_0$ as a function of the ratio of dynamical mass to 
 disc mass. The open squares correspond to the subsample of bulges with 
 $n \geqslant 2$, the filled squares refer to the subsample of bulges 
 with $n < 2$. Specially marked points, which deviate from the main relation, 
 refer to the galaxies with peculiarities in their structure 
 (see \cite{mos}).}}
 \label{DHCorr}
 \end{figure*}
\end{center}

We chose the ratio $\mathrm{M}_\mathrm{tot}/\mathrm{M}_\mathrm{d}$ 
to demonstrate a trend for the disc flattening with the relative mass of a 
spherical component (Fig.~\ref{DHCorr}). We adopted mass-to-light ratios 
$f_J = 1.5$, $f_H = 1.0$, and $f_{K_s} = 0.8$ $\mathrm{M}_\odot / L_\odot$ 
\cite{mcg} to get an estimate of the disc mass $\mathrm{M}_\mathrm{d}$. 
The scatter of points in Fig.~\ref{DHCorr} is rather large in contrast 
to the results displayed in \cite{zas02}. It may be caused by uncertainties 
in velocities and mass-to-light ratio determinations, decomposition errors 
and disc relaxation processes, that can thicken a disc just above the 
threshold for marginal stability. But some of this scatter must be real. 
Sotnikova and Rodionov \cite{sot05} concluded that the presence of a compact 
bulge is enough to suppress the bending instability that makes the disc 
thickness increase. A series of $N$-body simulations with the same total 
mass of a spherical component (dark halo $+$ bulge) were performed. 
The final disc thickness was found to be much smaller 
in the simulations, where a dense bulge is present, than in the simulations 
with bulgeless systems. The results of $N$-body simulations of discs starting 
from an unstable state were summarized in \cite{sot06}. The ratio 
$z_0 / h$ versus 
$(\mathrm{M}_\mathrm{h} + \mathrm{M}_\mathrm{b})/\mathrm{M}_\mathrm{d}$ 
($\mathrm{M}_\mathrm{h}$ is the halo mass within the sphere of radius 
$4h$ and $\mathrm{M}_\mathrm{b}$ is the bulge mass)
was plotted and it was shown that there is a clear scatter in this relation, 
in spite of the same model mass for a spherical component 
($\mathrm{M}_\mathrm{h} + \mathrm{M}_\mathrm{b}$). 

\begin{center}
 \begin{figure*}
 \begin{center}
  \includegraphics[width=5.4cm, angle=-90]{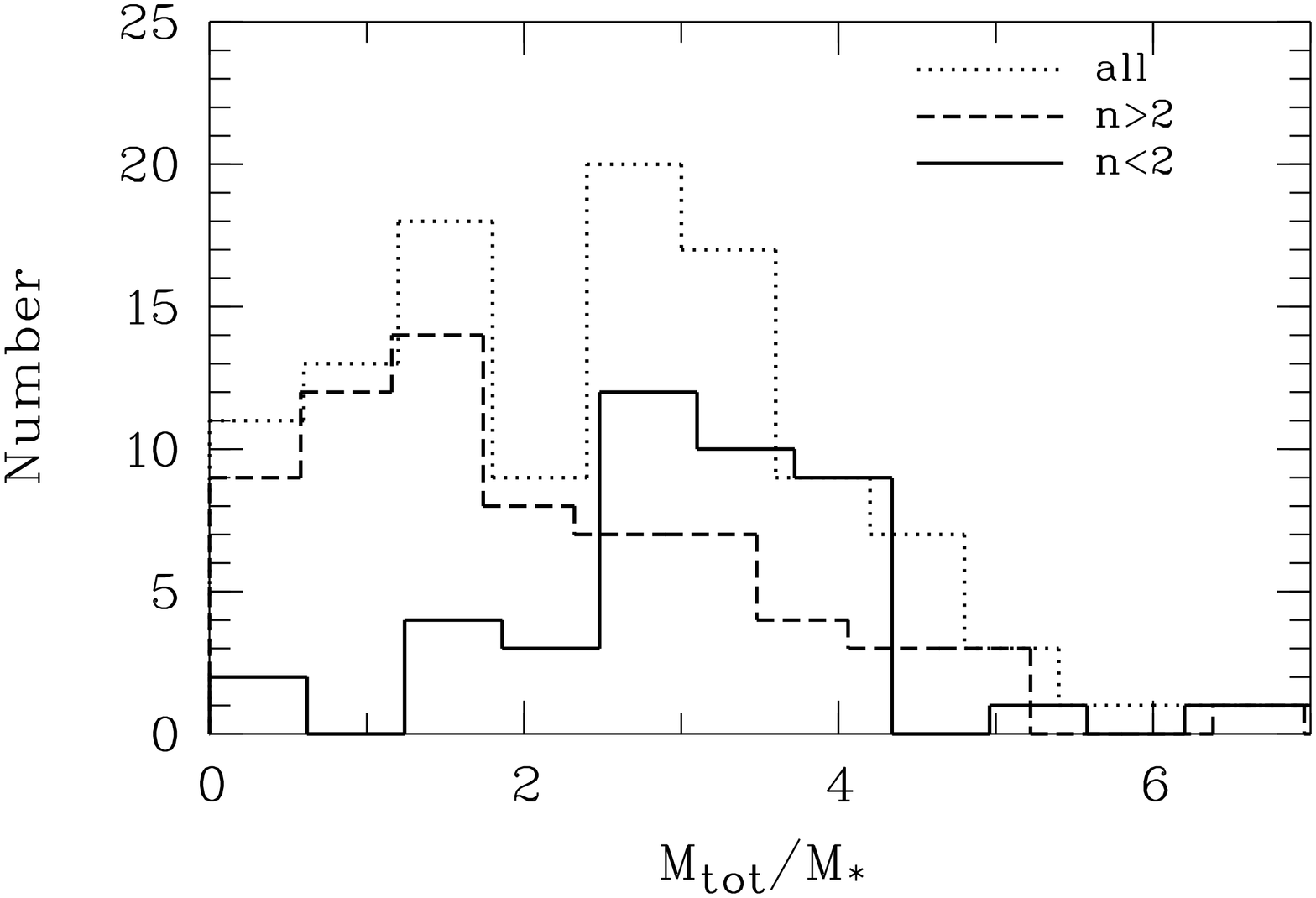}
  \includegraphics[width=5.4cm, angle=-90]{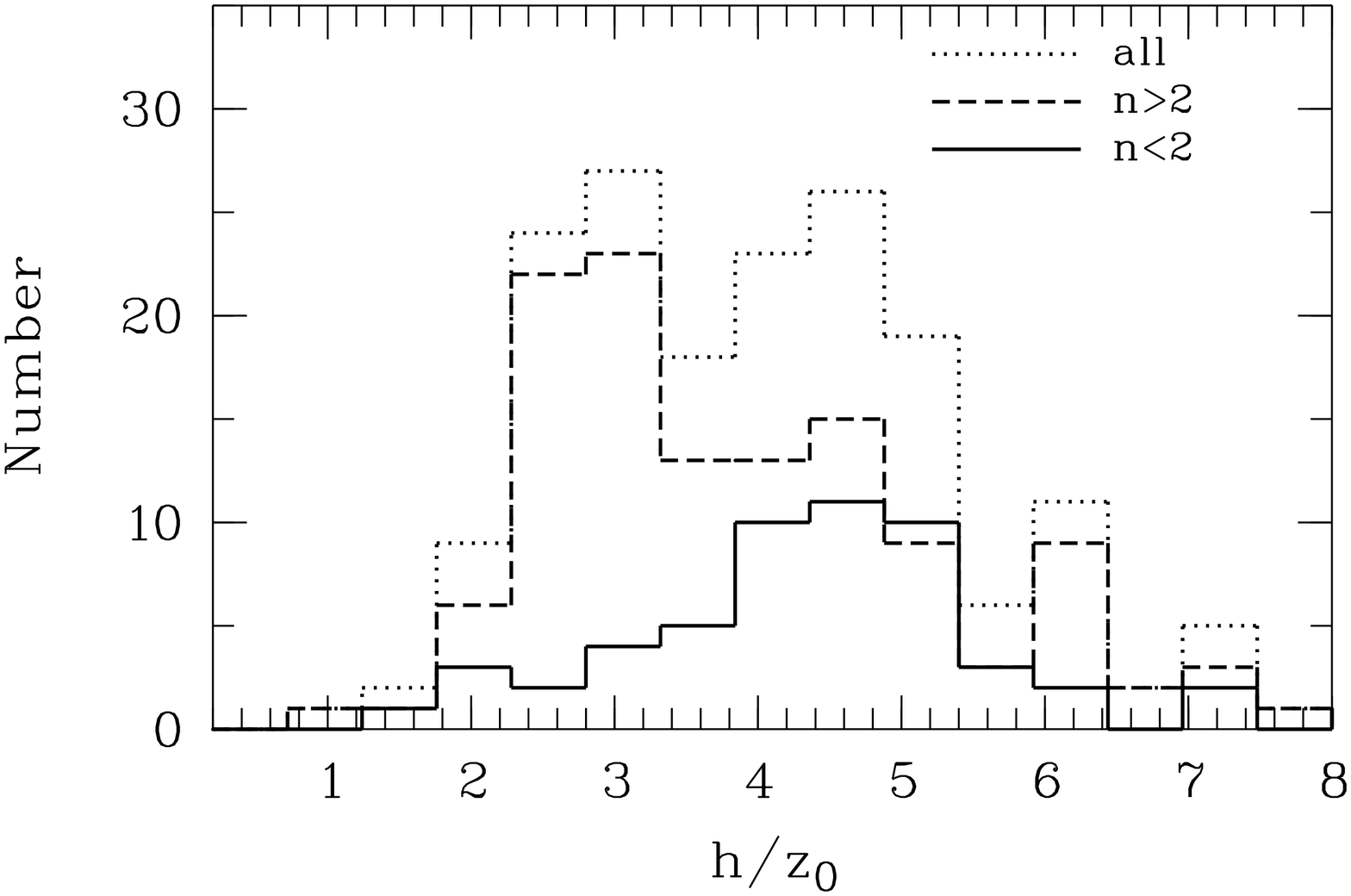}
 \end{center}
 \caption{\small{
 Distributions of the sample galaxies over the ratio of dynamical mass 
 to stellar mass of a galaxy (left) and the ratio of $h/z_0$ (right). 
 The dotted line indicates the distribution for the whole sample, 
 the dashed line corresponds to the subsample of bulges with 
 $n \geqslant 2$ and the solid line shows the distribution for 
 the subsample of bulges with $n < 2$.}}
 \label{DHDistrib}
 \end{figure*}
\end{center}

Fig.~\ref{DHCorr} shows no difference between galaxies with bulges and 
pseudobulges but there is a hint of bimodality in the distribution of our 
sample galaxies over the ratio of the dynamical mass to the stellar mass 
$\mathrm{M}_{*} = \mathrm{M}_\mathrm{d} + \mathrm{M}_\mathrm{b}$ 
(Fig.~\ref{DHDistrib}, the left plot) and over the ratio $h/z_0$ 
(Fig.~\ref{DHDistrib}, the right plot). This may reflect the presence 
of two different families of galaxies with different bulges in our sample. 
Galaxies with pseudobulges posess the large fraction of a dark 
component and their discs are relatively thiner.

\section[]{Summary and conclusions}

We constructed the large sample of edge-on spiral galaxies with  performed
2D bulge/disc decomposition of 2MASS galaxies images in $J$-, $H$- and
$K_s$-passband using the BUDDA v2.1 package. We extracted global bulge and
disc parameters for all objects. 

Analysing correlations between structural parameters, including those 
that describe the vertical structure, we found new evidences in favour of 
existence of two quite different populations of 
bulges in spiral galaxies --- classical bulges and pseudobulges.
Our results can be summarized as follows.
\begin{enumerate}
\item The bulge flattening $q_\mathrm{b}$ divides 
the sample into two different families --- triaxial, nearly prolate bulges 
and close to oblate bulges with moderate flattening. The \ser\ index 
threshold $n \simeq 2$ can be used to identify these two bulge types. 

\item The most surprising result arises from the investigation of the 
Photometric Plane of sample bulges. The bulges with $n\geqslant 2$ populate a
narrow strip in their Photometric Plane. However, there is a difference 
in behavior of this plane for bulges with $n\geqslant 2$ and $n < 2$. 
The plane is not flat and has a prominent curvature towards small values of 
$n$. This result may be due to the physical distinction between classical 
bulges and pseudobulges.

\item There is a correlation between the relative thickness of stellar discs 
$h/z_0$ and the relative mass of a spherical component, including a dark halo. 
This correlation was known previously for bulgeless galaxies 
\cite{zas02} or samples with predominantly bulgeless galaxies 
\cite{kre05} and was argued to arise from marginally stable discs. 
Our sample is much larger than the samples in \cite{kre05, zas02} 
and more reliable for statistical analysis. Here, 
we confirm the correlation under discussion. What is more, we did this 
not for bulgeless galaxies but for galaxies with massive bulges and concluded
that the {\it total} mass contained in spherical components 
may be one of the factors that determines the final steady state disc thickness.
\end{enumerate}

\section*{Acknowledgments}

This work was supported by the Russian Foundation for Basic Research 
(grant 09-02-00968).

\label{lastpage}

\end{document}